\documentclass[a4paper, 11pt]{article}

\usepackage{cite}
\usepackage{amsmath,amssymb,amsfonts}
\usepackage{algorithmic}
\usepackage{graphicx}
\usepackage{textcomp}
\usepackage{enumerate}
\usepackage{booktabs}
\usepackage{multirow}
\usepackage{caption}
\begin{document}
	\captionsetup[figure]{labelfont={bf},labelformat={default},labelsep=period,name={Fig.}}
	\captionsetup[table]{labelfont={bf},labelformat={default},labelsep=period,name={Table}}
	
	%\title{GAT-LSTM: A Robust Link Prediction Model for Temporal and Directed Networks}
	%Temporal generative and adversarial network with self-attention
	\title{TSAM: Temporal Link Prediction in Directed Networks based on Self-Attention Mechanism}

	\author{\small{Jinsong Li, Jianhua Peng, Shuxin Liu\footnote{Corresponding author. }, Lintianran Weng and Cong Li} \\ \small{\textit{PLA Strategic Support Force Information Engineering University}} \\ \small{Zhengzhou 450001, P. R. China} \\ \small{Email: liushuxin11@126.com}}
	
	\maketitle
	
	\begin{abstract}
		The development of graph neural networks (GCN) makes it possible to learn structural features from evolving complex networks. Even though a wide range of realistic networks are directed ones, few existing works investigated the properties of directed and temporal networks. In this paper, we address the problem of temporal link prediction in directed networks and propose a deep learning model based on GCN and self-attention mechanism, namely TSAM. The proposed model adopts an autoencoder architecture, which utilizes graph attentional layers to capture the structural feature of neighborhood nodes, as well as a set of graph convolutional layers to capture motif features. A graph recurrent unit layer with self-attention is utilized to learn temporal variations in the snapshot sequence. We run comparative experiments on four realistic networks to validate the effectiveness of  TSAM. Experimental results show that TSAM outperforms most benchmarks under two evaluation metrics. 
	\end{abstract}

%	\keywords{directed network, temporal link prediction, graph neural network, autoencoder, self-attention mechanism}
	
	\section{Introduction}
	Complex systems in real world can be naturally described with complex networks, where nodes represent entities and links represent the interactions between them. Complex networks are highly dynamic objects whose topology evolves quick over time with the appearance of new interactions~\cite{kumar_Structure_2006}. Predicting the dynamics of complex networks is a meaningful and promising problem. For example, in data center networks, the prediction of network topology can guide the design of routing protocols to improve the efficiency~\cite{lei_GCNGAN_2019}. In online social networks such as Twitter and Sina Weibo, the prediction of people's interactions can help infer the potential friends and recommend them to users, increasing their loyalty to the platform in return~\cite{Lu_LinkPrediction_2011}. In most literatures, such task is referred to as temporal link prediction, the goal of which is to predict future topology of evolving networks based on historic network information. Generally, temporal link prediction is more challenging than link prediction in static networks, because it requires to capture not only the structural feature of but also the temporal evolution~\cite{Tylenda_Timeaware_2009}. 
	
	A number of methods have been proposed to solve the temporal link prediction problem in the last two decades. Most existing works compact the historic network structures into one single network, and use methods in static networks to predict future links. Similarity-based temporal link prediction methods are the simplest and most efficient ones, which assume nodes with higher similarity will form links in the future~\cite{Liben_Linkprediction_2007}. Such methods include Common Neighbors, Jaccard, Adamic-Adar, Resource allocation, Katz,  etc~\cite{Lu_LinkPrediction_2011}. Even though these methods are efficient, they only take into account the structural feature of previous moment, regardless of informative temporal features such as the evolving pattern. Recently, the development of machine learning and graph neural network (GNN) makes it possible to build learning models for non-Euclidean data including complex networks~\cite{Kipf_Semisupervised_2017}. Taking advantage of GNN, many works focus on solving temporal link prediction problem with machine learning models which learn both structural and temporal feature at the same time. These works mainly fall into two categories. The first kind of methods use dynamic graph representation learning to learn latent representations of nodes in temporal networks, and train a downstream logistic regression classifier for link prediction~\cite{sankar_DySAT_2020}. The other kind of methods reconstruct the predicted network through an autoencoder architecture~\cite{lei_GCNGAN_2019}. Even though the latter one is more complicated with more parameters to train, it usually performs better in the task of temporal link prediction. 
	
	In reality, plenty of realist networks are directed in which link orientations have specified meanings. The prediction of link orientations are equally important with connectivity~\cite{Shang_RoleDirect_2017}. For example, in email networks, a directed link represents an outgoing email from one to another. If only the existence of a link is predicted, it would lead to ambiguity because one cannot decide the source and target of this email. Directed networks are quite different from undirected ones in topological properties, making temporal link prediction for directed networks a complicated and challenging problem. Since most existing works idealize their subjects into undirected networks, few works have investigated this problem thoroughly.  
	% Contributions and highlights:
	
	In this paper, we address the problem of predicting temporal links in directed networks. Our main contributions can be summarized as follows: 
	\begin{enumerate}[1.]
		\setlength{\itemsep}{0pt}\setlength{\parsep}{0pt}\setlength{\parskip}{0pt}\setlength{\topsep}{0pt}\setlength{\partopsep}{0pt}
		\item We design a model for link prediction in directed and temporal networks, namely TSAM. The model utilizes graph attentional layers to capture structural features of each snapshot. It also leverages matrix transformation to mine additional structural feature from local network structure. 
		\item To capture the temporal features efficiently and overcome the long term dependency of evolving structure, we use graph recurrent units with self-attention mechanism to learn from a sequence of snapshots and predict future snapshots. 
		\item Both node-level self-attention and time-level self-attention mechanisms are adopted in our model to accelerate the learning process and improve the prediction performance. 
		\item We use comparative experiments on realistic networks to validate the effectiveness of our model. 
	\end{enumerate}
	
	The remainder of this paper is organized as follows: Section~\ref{sec:relatedwork} introduces several related works. Section~\ref{sec:problemdes} describes the aiming problem of this paper. Section~\ref{sec:method} presents the proposed method. Section~\ref{sec:exps} describes experimental setups and analyzes the results. Finally, Section~\ref{sec:conclusion} draws conclusion of the paper. 
	
	\section{Related Works}\label{sec:relatedwork}
	\subsection{Temporal link prediction}
	Temporal networks are usually described in two ways: snapshot sequence which is a set of evolving snapshots at discrete time, and timestamped graph which is a graph with timestamped links. In this paper we adopt the first description. 
	
	Zhou et al.~\cite{Zhou_Dynamic_2018} leverage the concept of triadic closure as guidance to capture the evolving pattern across different snapshots. Goyal et al.~\cite{Goyal_DynGEM_2018} proposed DynGEM based on depth autoencoder to incrementally update node embeddings through initialization from the previous step. These methods cannot capture the dynamics among a long period of time, which leads to limit on accuracy. Since recurrent neural network (RNN) and its variations are powerful tools to capture temporal dynamics of input sequences, they are adopted in many temporal link prediction methods. Chen et al.~\cite{Chen_GCLSTM_2018} proposed GC-LSTM which uses graph convolutional network (GCN) embedded long short term memory network (LSTM) to predict temporal links. Instead of learning structural and temporal feature separately, Pareja et al.~\cite{Pareja_EvolveGCN_2019} use the RNN to evolve the GNN so that the dynamic features are captured in the evolving network parameters. Nevertheless, these recurrent methods are inefficient in capturing the most relevant historical snapshots because they treat the effect of each snapshot equally. In our proposed TSAM model, we utilize self-attention mechanism to differentiate the influence of snapshots. 
	
	On the other hand, some works have studied the problem of link prediction in directed and temporal networks. Jawed et al.~\cite{Jawed_TimeFrame_2015} addressed time frame based link prediction problem in directed citation networks and proposed time frame-based score. B\"ut\"un et al.~\cite{Butun_Extension_2018} designed a measure by extending neighbor based measures as directional pattern based ones to consider the role of link directions. 
	
	\subsection{Graph neural network}
	Graph convolution is a key technique to perform machine learning on non-Euclidean data structure such as complex networks. It generalizes the standard definition of convolution over a regular grid topology to graph structure. Graph aggregators are basic building blocks of graph convolution methods. Most existing works on graph aggregators are based on either pooling over neighborhoods or the weighted sum of neighboring features. Typically, graph convolutions can be categorized into two types: spectral domain convolution and spacial domain convolution. The classic GCN designed by Kipf et al.~\cite{Kipf_Semisupervised_2017} employs spectral domain convolution by leveraging the decomposition of Laplace matrix. Since the Laplace matrix should be symmetric to perform decomposition, GCN cannot deal with directed networks whose adjacency matrices are asymmetric. Hamilton et al.~\cite{hamilton_Inductive_2017} extended graph convolutional methods through trainable neighbor aggregation functions, and proposed a spacial domain convolution method named GraphSAGE. Since GraphSAGE does not use decomposition of Laplace matrix, it can deal with directed networks. Other efforts have also been done to learn representations of directed networks, such as motif2vec~\cite{Dareddy_Motif2vec_2019}, DIAGRAM~\cite{Kefato_WhichWay_2020}, ATP~\cite{Sun_ATPDirected_2018}, MotifNet~\cite{Monti_MotifNet_2018}, etc. 
	
	\subsection{Self-attention mechanism}
	Neural attention networks use a subnetwork to compute the correlation weight of the elements in a set. Among the family of attention models, the multi-head attention model proves to be effective for machine translation tasks. Later it has been adopted as a graph aggregator to solve the node classification problem~\cite{Velickovic_GraphAttention_2017} and static link prediction problem~\cite{Gu_LinkPrediction_2019}, referred to as graph attention network (GAT). Each attention head sums the elements that are similar to the query vector in one representation subspace, which provides more modeling power naturally. Zhang et al.~\cite{Zhang_GaANGated_2018} further proposed gated attention networks (GaAN) which treats the effect of different attention head unequally. Leveraging self-attention mechanism, Sankar et al.~\cite{sankar_DySAT_2020} proposed the DySAT to learn representation of temporal networks. Our work differs from theirs in two aspects: 1) On structural level, our TSAM model target directed networks and leverage additional structural features through matrix transformation and feature fusion, while DySAT aims at undirected networks. 2) On temporal level, TSAM adopts gated recurrent unit (GRU) layer and self-attention layer to capture temporal features, while DySAT only uses self-attention layer which cannot distinguish between different positions of the input.  
	
	\section{Problem description}\label{sec:problemdes}
	A directed and temporal network can be described with a sequence of network snapshots $ G = \{G_1, G_2, \cdots, G_T\} $, where $T$ is the number of time steps, $G_t = G({V}_t, {E}_t)$ is the snapshot at time step $t$, with ${V}_t$ and ${E}_t$ being the set of nodes and links, respectively. For simplicity, we only investigate the evolution of links and assume all snapshots share the same set of nodes, i.e., $V$. The adjacency matrix at time step $t$ can then be denoted as $\mathbf{A}_t = [a_{ij}^t]_{N \times N}$, $N = |V|$ is the total number nodes. Since we focus on directed and unweighted networks, when link $e(i,j) \in {E}_t$, $a_{ij}^t=1$, otherwise $a_{ij}^t=0$. Notice that in directed networks, $a_{ij}^t \neq a_{ji}^t$. Denote $\mathbf{X} \in \mathbb{R}^{N \times F}$ as the feature matrix of nodes, where $F$ is the number of features in each node. 
	
	Given the adjacency matrices during $T$, i.e., $\{\mathbf{A}_{t-T}, \mathbf{A}_{t-T+1}, \cdots, \mathbf{A}_{t}\}$, temporal link prediction problem aims at learning a function $f(\cdot)$ which can predict the adjacency matrix $\mathbf{A}_{t+1}$ at time step $t+1$ based on the link formation history, denoted as:
	\begin{equation}\label{eq:LP}
	\hat{\mathbf{A}}_{t+1} = f(\mathbf{A}_{t-T}, \mathbf{A}_{t-T+1}, \cdots, \mathbf{A}_t)
	\end{equation}
	
	The goal of our model is to learn function $f(\cdot)$ of a given network and use it to predict its future links. For simplicity, in the following we denote $\mathbf{A}_{t-T}^t = \{\mathbf{A}_{t-T}, \mathbf{A}_{t-T+1}, \cdots, \mathbf{A}_{t}\}$. 
	
	%\begin{table}[htbp]
	%	\caption{Notations used in TSAM. }
	%	\footnotesize
	%	\renewcommand\arraystretch{1.2}
	%	\begin{center}
	%		\begin{tabular}{lr}
	%			\hline
	%			Symbol & Definition \\
	%			\hline
	%			$T$ & number of time steps\\
	%			$G_t$ & snapshot at time step $t$\\
	%			$N$ & number of nodes in each snapshot \\
	%			$\mathbf{A}_t \in {\mathbb{R}^{N \times N}}$ & adjacency matrix of snapshot $t$ \\
	%			${\cal{N}}^{\mathrm{in}}_t(u)$ & incoming neighbors of node $u$ in snapshot $t$ \\
	%			$F$ & dimension of node features\\
	%			$\mathbf{X} \in {\mathbb{R}}^{N \times F}$ & feature matrix in each snapshot\\
	%			$\cdot^{\mathrm{T}}$ & vector transposition operation\\
	%			$||$ & vector concatenation operation\\
	%			$K$ & number of attention head in GAT layer\\
	%			\hline
	%		\end{tabular}%
	%		\label{tab:notation}
	%	\end{center}
	%\end{table}
	
	\section{The Proposed Method}\label{sec:method}
\begin{figure}[h]
		\centerline{\includegraphics[width=14cm]{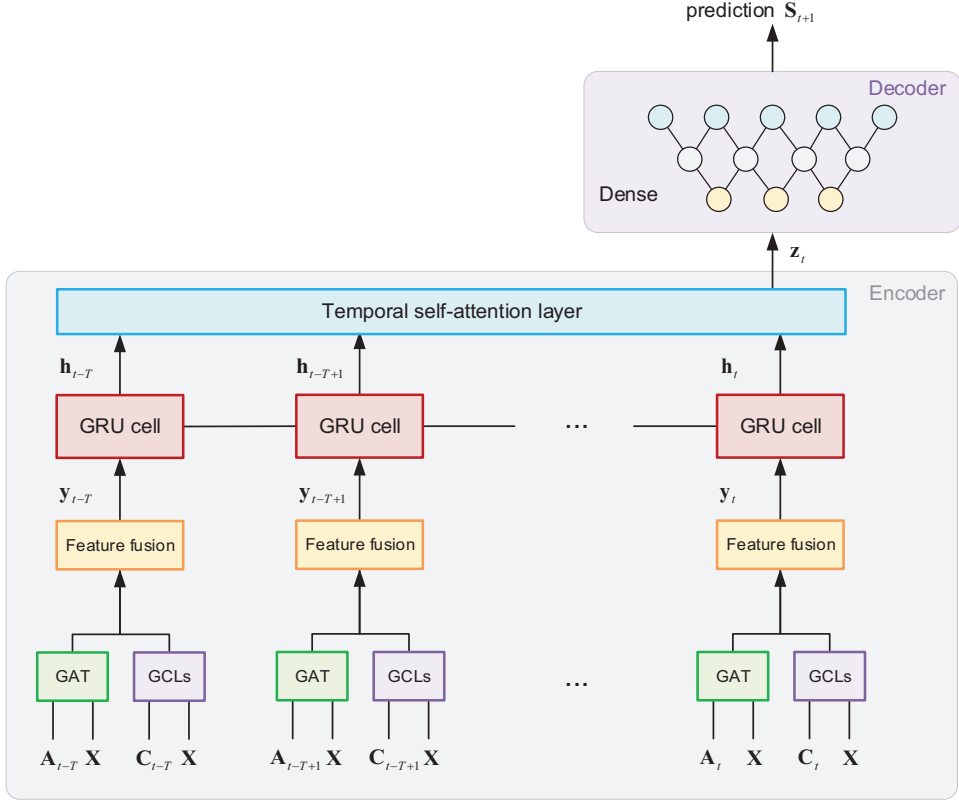}}
		\caption{Overall architecture of the proposed TSAM model. }
		\label{fig:model}
	\end{figure}

	%\subsection{Model architecture}
	We propose a temporal link prediction model based on self-attention mechanism, referred to as TSAM. The basic architecture of TSAM model is an autoencoder as shown in Fig.~\ref{fig:model}. First, the temporal encoder consisted of graph convolutional layers, graph attention layers and GRU layers draws both the structure-related and time-related features from the input snapshots, generating node embedding of the last time step. Then the decoder utilizes full-connected layers to interpret the embedding to the predicted adjacency matrix. Here we introduce the main parts of TSAM model separately. 
	
	\subsection{Node-level attention block}
	For each snapshot at time step $t$, we take advantage of the graph attentional (GAT) layer to specify different weights to different nodes in a neighborhood. The GAT layer attends over the immediate neighbors of a node in snapshot $G$, by computing the attention weights as a function of their input feature vectors. At time step $t$, the inputs to the GAT layer are the feature matrix $\mathbf{X} = \{\mathbf{x}_1, \mathbf{x}_2, \cdots, \mathbf{x}_N\}$, $\mathbf{x}_i \in \mathbb{R}^{F}$ and the adjacency matrix $\mathbf{A}_t \in \mathbb{R}^{N \times N}$, where $N$ is the number of nodes, $F$ is the number of features in each node. The output of GAT layer is a new set of features $\mathbf{Y}^o = \{\mathbf{y}^o_1, \mathbf{y}^o_2, \cdots, \mathbf{y}^o_N\}$, $\mathbf{y}_i \in \mathbb{R}^{F'}$, where $F'$ is the dimension of new features in each node. First, a shared linear transformation parametrized by $\mathbf{W}^{(n)} \in \mathbb{R}^{F' \times F}$ is applied to every node. Then a shared attentional mechanism $a: \mathbb{R}^{F'} \times \mathbb{R}^{F'} \to \mathbb{R}$ computes the attention coefficients of two nodes as:
	\begin{equation}\label{eq:nodeattn1}
	e_{ij} = LeakyReLU\left(\mathbf{a}^\mathrm{T} \cdot Concat(\mathbf{W}^{(n)} \mathbf{x}_i, \mathbf{W}^{(n)} \mathbf{x}_j)\right)
	\end{equation}
	where $LeakyReLU(\cdot)$ is the LeakyReLU activation function with $\alpha=0.2$, $\cdot^\mathrm{T}$ represents vector transposition operation, $Concat(\cdot, \cdot)$ represents vector concatenation operation. 
	
	The calculated attention coefficient $e_{ij}$ represents the importance of node $i$ to node $j$. Since we focus on directed networks, we perform the so-called masked attention which only computes $e_{ij}$ for nodes $j \in {\cal{N}}^{\mathrm{in}}_t(i)$, where ${\cal{N}}^{\mathrm{in}}_t(i)$ is the first-order incoming neighbors of node $i$ in the snapshot. The attention coefficients are then normalized with the softmax function, denoted as:
	\begin{equation}\label{eq:nodeattn2}
	\alpha_{ij} = \frac{\mathrm{exp}(e_{ij})}{\sum_{k\in {\cal{N}}^{\mathrm{in}}_t(i)} \mathrm{exp}(e_{ik})}
	\end{equation}
	
	We follow~\cite{Velickovic_GraphAttention_2017} to perform the multi-head attention mechanism which can stabilize the learning process of self-attention. $K_N$ independent attention mechanisms are performed with the output features concatenated as the final result, denoted as: 
	\begin{equation}\label{eq:nodeattn3}
	\mathbf{y}^o_i = ELU\left(\frac{1}{K_N}\sum_{k = 1}^{{K_N}} \sum_{j\in {\cal{N}}^{\mathrm{in}}_t(i)} \alpha^k_{ij} \mathbf{W}^{(n)}_k \mathbf{x}_i\right)
	\end{equation}
	where $ELU(\cdot)$ is the exponential linear unit (ELU) activation function. 
	
	\subsection{Feature generation}\label{feature}
	In order to leverage more structural information of directed networks, we generate a set of transformed adjacency matrices using the simplest matrix operations. In detail, at each time step $t$, we define a set of mapping functions $\{g_{M_1}, g_{M_2}, \cdots, g_{M_i}\}$ where $g:{\mathbb{R}}^{N \times N} \to {\mathbb{R}}^{N \times N}$ maps the adjacency matrix to the transformed adjacency matrices as: 
	\begin{equation}\label{eq:transform}
	{\mathbf{C}}_t^{M_i} = g_{M_i}({\mathbf{A}}_t)
	\end{equation}
	
	There are multiple choices of mapping functions. Here we list four simplest forms as:
	\begin{equation}\label{eq:transexp}
	{\mathbf{C}}_t^{M_1} = {\mathbf{A}} \cdot {\mathbf{A}}, 
	{\mathbf{C}}_t^{M_2} = {\mathbf{A}}^\mathrm{T} \cdot {\mathbf{A}}, 
	{\mathbf{C}}_t^{M_3} = {\mathbf{A}} \cdot {\mathbf{A}}^\mathrm{T}, 
	{\mathbf{C}}_t^{M_4} = {\mathbf{A}}^\mathrm{T} \cdot {\mathbf{A}}^\mathrm{T}
	\end{equation}
	
\begin{figure}[h]
		\centerline{\includegraphics[width=11 cm]{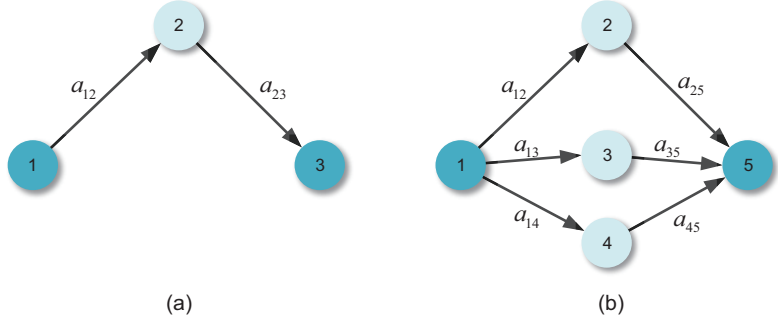}}
		\caption{Example of matrix transformation in directed networks. }
		\label{fig:example}
	\end{figure}

	The meaning of these four transformed adjacency matrices can be interpreted with network motifs~\cite{Bianconi_LocalStructure_2008}. Take two simple networks shown in Fig.~\ref{fig:example} as example. In Fig.~\ref{fig:example}(a), operation $\mathbf{A} \cdot \mathbf{A}$ can be calculated as: 
	\begin{equation}\label{eq:example1}
	{\left( {{\bf{A}} \cdot {\bf{A}}} \right)_{13}} = \sum\limits_{k = 1}^3 {{a_{1k}} \cdot {a_{k3}} = {a_{12}} \cdot {a_{23}}} = 1
	\end{equation}
	
	In Fig.~\ref{fig:example}(b), operation $\mathbf{A} \cdot \mathbf{A}$ can be calculated as: 
	\begin{equation}\label{eq:example2}
	{\left( {{\bf{A}} \cdot {\bf{A}}} \right)_{15}} = \sum\limits_{k = 1}^5 {{a_{1k}} \cdot {a_{k3}} = } {a_{12}} \cdot {a_{25}} + {a_{13}} \cdot {a_{35}} + {a_{14}} \cdot {a_{45}} = 3
	\end{equation}
	
	Obviously, operation $\mathbf{A} \cdot \mathbf{A}$ counts the number of motif $\{u \to t \to v\}$ between two nodes, generating a symmetric matrix whose elements stand for the number of such type of motif. Other operations in~(\ref{eq:transexp})  have similar meanings. 
	
	We use a set of graph convolutional layers (GCL) to exact structural features and generate corresponding embeddings from the transformed adjacency matrices. For each transformed adjacency matrix $\mathbf{C}^{M_i}_t$, the output of GCL can be denoted as: 
	\begin{equation}\label{eq:gcl1}
	\mathbf{Y}^{M_i}_t = ELU\left(\hat{\mathbf{D}}^{-1/2} \hat{\mathbf{C}}^{M_i}_t \hat{\mathbf{D}}^{-1/2} \mathbf{X} \mathbf{W}^{(g)}\right)
	\end{equation}
	where $\hat{\mathbf{C}}^{M_i}_t = \mathbf{C}^{M_i}_t + \mathbf{I}_N$, $\mathbf{I}_N$ is the $N$-dimensional identity matrix. $\hat{\mathbf{D}}_{uu} = \sum_{v=1}^{N} (\hat{\mathbf{C}}^{M_i}_t)_{uv}$ is the degree matrix. 
	
	The extracted features from GAT layer and the set of GCLs captures different structural properties of the input network $G_t$. We adopt a feature fusion layer to combine them together. For such early fusion tasks, concatenation method and addition method are two simplest and effective ways. Here we adopt the element-wise addition method to ensure that the output feature has the same dimension with the input. Notice that possible information losses may occur when features are added up directly, but computational costs are saved at the same time. The output of feature fusion layer is denoted as: 
	\begin{equation}\label{eq:gcl2}
	\mathbf{Y} = LN\left(Add(\mathbf{Y}^o, \mathbf{Y}^{M_1}, \cdots, \mathbf{Y}^{M_i})\right)
	\end{equation}
	where $LN(\cdot)$ is the layer normalization function to normalize the outputs with range $[0,1]$, $Add(\cdot, \cdot)$ is the element-wise add function. Finally, a flatten layer reshapes the output embeddings into row-wise vectors in order to feed them into the RNN networks later. 
	
	\begin{figure}[h]
		\centerline{\includegraphics[width=11 cm]{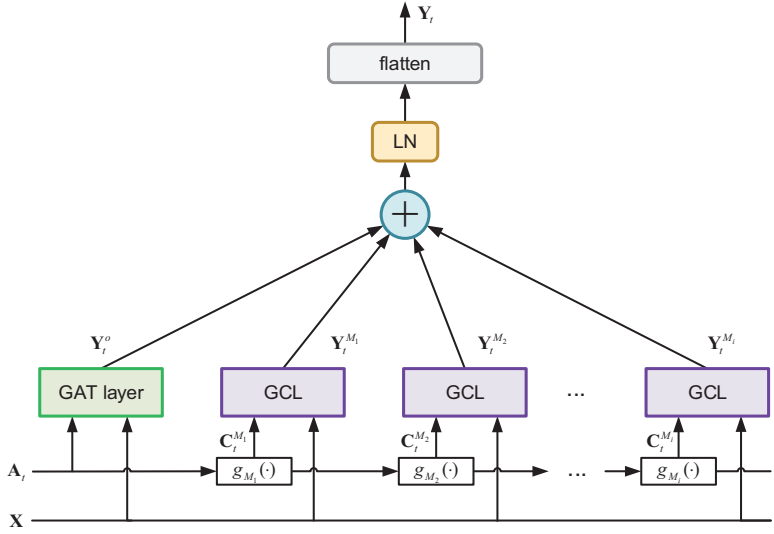}}
		\caption{Diagram of feature generation in each snapshot. }
		\label{fig:feature}
	\end{figure}
	
	\subsection{Time-level attention block}
	 To capture the temporal variations of network structure through multiple time steps, we feed the sequence of row-wise vectors into a time-level attention block. The time-level attention block consists of two layers: a nonlinear RNN layer and a temporal self-attention layer.
	 
	Recurrent neural networks has flexible nonlinear transformation ability on time-series inputs, while the attention mechanism has limited representational power for it uses weighted sum to generate output vectors. To increase the expressive power on time-level, the row-wise vector sequence $ \mathbf{Y}^t_{t-T} = \{\mathbf{y}_{t-T}, \mathbf{y}_{t-T+1}, \cdots, \mathbf{y}_t\}, \mathbf{y}_t \in \mathbb{R}^{1 \times (N \times F')}$ are first fed into a recurrent neural network to mine the evolving patterns of the temporal directed network. LSTM and GRU are two well-performing models in RNN networks which are capable to learn the long-term dependencies of sequential data. Since GRU is able to achieve similar performance compared with LSTM with fewer trainable parameters and lower computational complexity, we use GRU hidden layer to deal with this specific task. At each time step $t$, the input vector $\mathbf{y}_t$ and the last time step state vector $\mathbf{h}_{t-1}$ are taken as the input of the GRU cell, then the output state vector $\mathbf{h}_t$ can be denoted as: 
	\begin{equation}\label{eq:gru1}
	\mathbf{a}_t = \sigma\left(\mathbf{W}^{(z)} \mathbf{y}_t + \mathbf{U}^{(z)} \mathbf{h}_{t-1} + \mathbf{b}^{(z)}\right)
	\end{equation}
	\begin{equation}\label{eq:gru2}
	\mathbf{r}_t = \sigma\left(\mathbf{W}^{(r)} \mathbf{y}_t + \mathbf{U}^{(r)} \mathbf{h}_{t-1} + \mathbf{b}^{(r)}\right)
	\end{equation}
	\begin{equation}\label{eq:gru3}
	\tilde{\mathbf{h}}_t = \mathrm{tanh}\left(\mathbf{W}^{(n)} \mathbf{y}_t + \mathbf{r}_t \odot \mathbf{U}^{(n)} + \mathbf{b}^{(n)}\right)
	\end{equation}
	\begin{equation}\label{eq:gru4}
	\mathbf{h}_t = \left(1-\mathbf{z}_t\right) \odot \mathbf{h}_{t-1} + \mathbf{z}_t \odot \tilde{\mathbf{h}}_t
	\end{equation}
	where $ \{\mathbf{W}^{(i)}, \mathbf{U}^{(i)}, \mathbf{b}^{(i)}\}, i=\{z, r, n\}$ are the trainable parameters of the update gate, reset gate and new memory, respectively. $\sigma(\cdot)$ and $\mathrm{tanh}(\cdot)$ respectively represent the sigmoid and tanh activation function. $\odot$ denotes the Hadamard product. We denote $H_R$ as the hidden layer dimension of the GRU layer. 
	
\begin{figure}[t]
		\centerline{\includegraphics[width=12 cm]{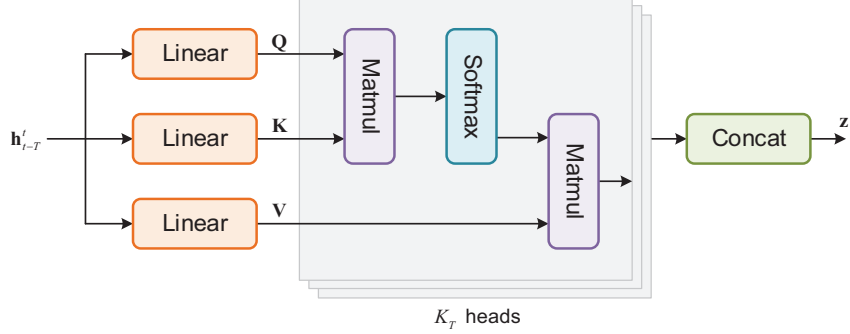}}
		\caption{Diagram of temporal self-attention layer. }
		\label{fig:tsa}
	\end{figure}

	Afterwards, the hidden states of GRU layer are fed into a temporal self-attention layer to differentiate their influences on each other. In detail, we take the hidden state vector at time step $t$, i.e., $\mathbf{h}_t$, as the query to attend over its historical representations, tracing the evolution of structural features. We follow~\cite{vaswani_Attention_2017} to adopt scaled dot-product attention in order to accelerate computational speed, as shown in Fig.~\ref{fig:tsa}. At each time step, the temporal self-attention layer takes the hidden state vectors $\mathbf{h}_{t-T}^t = \{ {{\mathbf{h}}_{t - T}},{{\bf{h}}_{t - T + 1}}, \cdots ,{{\mathbf{h}}_t}\} $ as input and produces a new sequence $\mathbf{z}_{t-T}^t = \{ {{\mathbf{z}}_{t - T}},{{\mathbf{z}}_{t - T + 1}}, \cdots ,{{\mathbf{z}}_t}\} $. Similar with node-level attention block, we employ multi-head attention mechanism to learn features from different latent space and enhance the representational ability of our model~\cite{voita_Analyzing_2019}. In the $l$-th attention head, linear transformations are performed first on the input vector to generate the queries, keys and values, denoted as:
	\begin{equation}\label{eq:tattn1}
	\mathbf{Q} =\mathbf{h}_{t-T}^t \mathbf{W}^{(q)}, \mathbf{K} =\mathbf{h}_{t-T}^t \mathbf{W}^{(k)}, \mathbf{V} =\mathbf{h}_{t-T}^t \mathbf{W}^{(v)}
	\end{equation} 
	where $\mathbf{W}^{(q)}, \mathbf{W}^{(k)}, \mathbf{W}^{(v)} \in \mathbb{R}^{H_R \times F''}$ are the trainable weights, $F''$ is the output feature dimension. Then the output vector is computed as:
	\begin{equation}\label{eq:tattn2}
	\mathbf{z}^{(l)} = \boldsymbol{\beta} \mathbf{V}
	\end{equation}
	where $\beta_{ij} \in \boldsymbol{\beta}$, 
	\begin{equation}\label{eq:tattn3}
	\mathbf{\beta}_{ij} = \frac{{\exp (e_{ij})}}{{\sum\limits_{k = 1}^T {\exp (e_{ik})} }}, i,j \in [t-T, t]
	\end{equation}
	is the softmax function, and 
	\begin{equation}\label{eq:tattn4}
	\mathbf{e} = \frac{\mathbf{Q} \mathbf{K}^\mathrm{T}}{\sqrt{F''}}
	\end{equation}
	is the attention coefficient matrix, with $e_{ij}\in\mathbf{e}$ indicating the influence of snapshot $i$ on snapshot $j$. Since the input vectors are time-relevant, we follow~\cite{sankar_DySAT_2020} to add a mask matrix $\mathbf{M} \in \mathbb{R}^{l \times l}$ in~(\ref{eq:tattn4}) to enhance the auto-regressive property, such that: 
	\begin{equation}\label{eq:tattn5}
	e^{ij} = \frac{\left(\mathbf{Q} \mathbf{K}^{\mathrm{T}}\right)_{ij}}{{\sqrt {F''} }} + {{M}_{ij}}
	\end{equation}
	where ${M}_{ij}=-\infty$ when $i>j$, otherwise $0$. This makes sure that when $i>j$, the softmax operation in~(\ref{eq:tattn2}) generates a zero attention weight, i.e., $\beta_{ij}=0$, which can ignore the attention from time step $i$ to $j$. 
	The outputs of $K_T$ independent attention heads are concatenated together as the final embedding, denoted as:
	\begin{equation}\label{eq:tattn6}
	\mathbf{z}_{t-T}^t = Concat\left(\mathbf{z}^{(1)}, \mathbf{z}^{(2)}, \cdots, \mathbf{z}^{(K_T)}\right)
	\end{equation}
	
	In this case, the shape of $\mathbf{z}_{t-T}^t$ is $T \times (K_T \times F'')$. 
	\subsection{The decoder network}
	In order to predict the network at time $t+1$, we treat the output embedding at time $t$, i.e. $\mathbf{z}_{t} \in \mathbb{R}^{1 \times (K_T \times F'')}$ as the learned embedding of the historical snapshots. The output vector is then fed into a decoder network to generate the prediction result. Briefly, the decoder network consists a fully-connected layer, denoted as: 
	\begin{equation}\label{eq:ffn}
	D(\mathbf{z}_t) = ReLU\left(ReLU\left({\mathbf{z}_t} {\mathbf{W}^{(h)}} + {\mathbf{b}^{(h)}}\right) {\mathbf{W}^{(o)}} + {\mathbf{b}^{(o)}}\right)
	\end{equation}
	where $\mathbf{W}^{(h)} \in \mathbb{R}^{(K_T \times F'') \times H_D}$ and $\mathbf{W}^{(o)} \in \mathbb{R}^{H_D \times (N \times N)}$ are the weights of the hidden layer and output layer, $H_D$ is the dimension of hidden layer. We denote $\mathbf{S}_{t+1}=D(\mathbf{z}_t)$ as the output vector of the decoder network, which is then reshaped from $1 \times (N \times N)$ to $N \times N$, with $[\mathbf{S}_{t+1}]_{ij} \in [0,1]$ representing the existence probability of links in snapshot $G_{t+1}$. 
	
	\subsection{Model optimization}
	Since our goal is to predict links in $G_{t+1}$ based on historical snapshots, the predicted score matrix $\mathbf{S}_{t+1}$ and the ground-truth adjacency matrix $\mathbf{A}_{t+1}$ should be close geometrically. In other words, if link $e(i,j) \in \mathbf{A}_{t+1}$, the corresponding score $s_{ij} \in \mathbf{S}_{t+1}$ is close to $1$ for a good prediction, otherwise $0$. We use Frobenius norm to describe the distance between matrices, and train the TSAM model at time step $t$ by optimizing the following loss function:
	\begin{equation}\label{eq:loss}
	{L_t}(\theta; {\mathbf{A}}_{t-T}^{t}, {\mathbf{A}}_{t+1}) = \left\| {\left({{{\mathbf{S}}}_{t + 1}} - {{\mathbf{A}}_{t + 1}}\right) \odot {\cal B}} \right\|_F^2 + \frac{\lambda}{2} \left\|\theta\right\|_2^2
	\end{equation}
	where ${\mathbf{S}}_{t+1}=f({\mathbf{A}}_{t-T}^t)$ is the predicted score. ${\cal B}$ is the penalty term to deal with the sparsity of adjacency matrix following literature~\cite{Wang_Structure_2016}. ${\cal B}_{ij}=\beta$ for $e(i,j) \in \mathbf{A}_{t+1}$, other ${\cal B}_{ij}=1$. In such case, when $\beta > 1$, it penalize inaccurate predictions of observed links more than those of unobserved links. To encourage sparsity in the model's weights and prevent over-fitting, we add ${\cal L}_2$ regularizer to the loss function with $\lambda$ as the hyperparameter controlling its relative weight. We adopt Adam optimizer to minimize the loss function. 
	
	\section{Experiments}\label{sec:exps}
	
	\subsection{Datasets}
	We use four temporal directed networks from real world to evaluate the performance of TSAM model, including two Email networks, a social network and an interaction network. Basic statistics of the four datasets are presented in Table~\ref{tab:dataset}, and a brief introduction of them is described as follows. 
	
	\begin{table}[htbp]
		\caption{Summary statistics of four temporal networks. }
		\footnotesize
		\renewcommand\arraystretch{1.2}
		\begin{center}
			\begin{tabular}{ccccc}
				\hline
				Network & MAN   & EEC   & UCI   & LEM \\
				\hline
				$\#$ of Nodes        & 167          & 964            & 889              & 485 \\
				$\#$ of Links          & 81,127     & 291,167      & 10,034          & 196,364 \\
				Average degree      & 971.58     & 604.08       & 22.57           & 809.75  \\
				Start date               & 2010/1/3   & xxxx/3/17  & 2004/6/27     & 1979/4/1 \\
				End date                & 2010/9/27 & xxxx/6/7    & 2004/10/26    & 2004/6/1 \\
				Total time span      & 9 months   & 15 months & 4 months       & 302 months \\
				Snapshot range      & 1week       & 1week      & 3 days          & 6 months \\
				$\#$ of Snapshots  & 38             & 64             & 40                & 50 \\
				window size $T$    & 8               & 8              & 5                 & 8   \\
				\hline
			\end{tabular}%
			\label{tab:dataset}
		\end{center}
	\end{table}
	
\begin{figure}[t]
		\centerline{\includegraphics[width=14 cm]{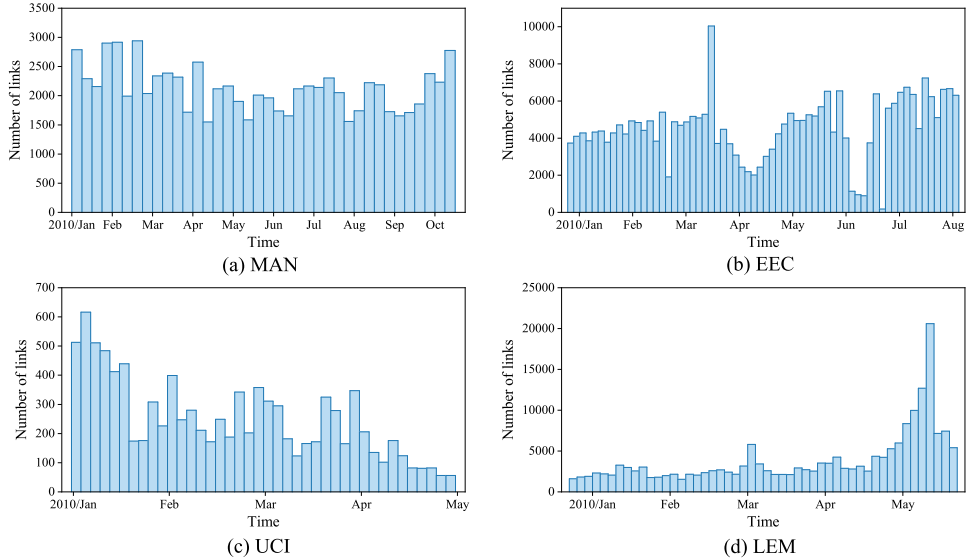}}
		\caption{Histogram of link numbers for each snapshot in four networks. }
		\label{fig:histplot}
	\end{figure}

	\begin{enumerate}[1.]
		\setlength{\itemsep}{0pt}\setlength{\parsep}{0pt}\setlength{\parskip}{0pt}\setlength{\topsep}{0pt}\setlength{\partopsep}{0pt}
		\item Manufacturing emails (MAN)~\cite{Kunegis_KONECT_2013}: An email network between employees of a mid-sized manufacturing company. The network is directed and nodes represent employees. The left node represents the sender and the right node represents the recipient. A directed link $e(u,v,t)$ represents that employee $u$ sent an email to employee $v$ at time $t$. We use links in one week to generate a snapshot, and construct $38$ snapshots with total duration of $9$ months, as shown in Fig.~\ref{fig:histplot}(a). 
		\item Email-Eu-core-temporal (EEC)~\cite{Paranjape_Motifs_2017}: An email network generated from the email data in a large European research institution. A directed link $e(u,v,t)$ represents that a person $u$ sent an email to another person $v$ at time $t$. The timestamp of this dataset starts from $0$ and no starting date is specified. We neglect the isolated links and only generate the snapshots based on links occurred during 15 consecutive months. Each snapshot contains links during one week, so we totally get $64$ snapshots as shown in Fig.~\ref{fig:histplot}(b). 
		\item UC Irvine messages (UCI)~\cite{Panzarasa_Patterns_2009}: A network consisted of private messages sent on an online social network at the University of California, Irvine. A directed link $e(u,v,t)$ represents that user $u$ sent a private message to user $v$ at time $t$. We choose links of 4 months in the experiments, and generate snapshots with the range of $3$ days as shown in Fig.~\ref{fig:histplot}(c). 
		\item LevantMonths (LEM)~\cite{Leskovec_SNAP_2014}: An interaction network collected by the Kansas Event Data System based on folders containing WEIS-coded events within eight countries: Egypt, Israel, Jordan, Lebanon, Palestinians, Syria, USA, and Russia. The dataset contains interactions from April 1979 to June 2004. Each snapshot contains links occurred in $6$ months, and $50$ snapshots are constructed in total, as shown in Fig.~\ref{fig:histplot}(d). 
	\end{enumerate}

	\subsection{Evaluation metrics}\label{evalmetric}
	We use two standard evaluation metrics to evaluate the performance of temporal link prediction models. 
	The area under the receiver characteristic operator curve (AUC) is a widely adopted evaluation metric for classification models, which considers both the sensitivity and specificity of the model. In link prediction problems, if there are $n'$ times that the scores of randomly chosen existent links are higher than those of randomly chosen non-existent links among $n$ independent comparisons, and $n''$ times that they get the same scores, then AUC is calculated as:
	\begin{equation}\label{eq:AUC}
	\mathrm{AUC} = \frac{n'+0.5n''}{n}
	\end{equation}
	
	A larger AUC score indicates a better prediction performance for a given model. Similar with AUC, the area under the precision-recall curve (PRAUC) is designed to evaluate the sparsity of networks. However, both AUC and PRAUC cannot evaluate the added and removed links at the same time. Therefore, in addition we adopt the geometric mean of AUC and PRAUC (GMAUC)~\cite{Junuthula_Evaluating_2016} to evaluate both added and removed links, defined as:
	\begin{equation}\label{eq:GMAUC}
	\mathrm{GMAUC} = \left(\frac{\mathrm{PRAUC}-\frac{L_A}{L_A + L_R}}{1-L_A/(L_A+L_R)} \cdot 2\left(\mathrm{AUC}-0.5\right)\right)^{1/2}
	\end{equation}
	where $L_A$ and $L_R$ are the number of added and removed links respectively, $\mathrm{PRAUC}$ is the $\mathrm{PRAUC}$ value of new links, while $\mathrm{AUC}$ is the AUC score calculated by originally existed links. 
	
	\subsection{Performance evaluation}\label{performance}
	
	Experiments are performed on a Ubuntu 16.04 LTS system with $48$ cores, $128$ GB RAM and $2.20$ GHz clock frequency. We implement the model in Tensorflow 1.15.0, and train it on a NVIDIA Titan Xp GPU. We use a sliding window with size $T$ to get continuous snapshot sequences from all snapshots in each network. At each time step $t$, we train separate models up to snapshot $t$ and evaluate it at $t+1$ for each $t=1, \cdots, T$. 
	
	\begin{table*}[h]
		\caption{Parameter settings of TSAM in four networks. }
		\footnotesize
		\renewcommand\arraystretch{1.2}
		\begin{center}
			\begin{tabular}{cccccccccc}
				\toprule
				Network & $N$     & $F'$   & $H_R$   & $F''$   & $K_N$    & $K_T$    & $H_D$  & $lr$  & $\lambda$\\
				\midrule
				MAN   & $167$   & $32$    & $1024$   & $256$   & $4$         &  $8$   & $128$  & $0.001$  & $0$\\
				EEC     & $964$   & $64$   & $4096$    & $1024$    & $2$         &  $4$    & $512$  & $0.001$  & $1e{-5}$\\
				UCI     & $889$   & $64$    & $4096$    & $1024$    & $2$         &  $4$    & $512$  & $0.001$  & $1e{-5}$\\
				LEM    & $485$   & $32$    & $2048$    & $512$    & $4$         &  $8$    & $256$  & $0.005$  & $0$\\
				\bottomrule
			\end{tabular}%
			\label{tab:parameter}
		\end{center}
	\end{table*}

	We compare the performance of TSAM and five state-of-the-art models on temporal link prediction in directed networks, including: 
	
	\begin{enumerate}[1.]
		\item TNE~\cite{Zhu_Scalable_2016}: It models the temporal network with Markov process and uses matrix factorization to learn node embeddings. 
		\item GC-LSTM~\cite{Chen_GCLSTM_2018}: It is an end-to-end temporal link prediction model which uses GCN to extract structural features and LSTM to extract temporal features. We set the order of GCN $K=3$ to aggregate 3-hop neighbors. 
%		\item GCN-GRU: It is a modified model originated from GCRN~\cite{Seo_Structured_2016}, a deep learning model for predicting structured sequences of data. It combines GCN on graphs to identify spatial structures and RNN to find dynamic patterns. We choose GRU in the RNN layer and use it to predict temporal links following the optimization method of TSAM. 
		\item EvolveGCN~\cite{Pareja_EvolveGCN_2019}: It is a graph embedding model which adapts GCN model along the temporal dimension without resorting to node embeddings. It captures the dynamism of graph sequences using an RNN to evolve the GCN parameters. 
		\item dyngraph2vec~\cite{Goyal_Dyngraph2vec_2018}: It uses multiple non-linear layers to learn structural patterns of each snapshot, and then uses recurrent layers to learn temporal transitions in the network. We choose one of its variations namely dyngraph2vetAERNN, which uses LSTM in the encoder to extract node embeddings and full-connected network in the decoder to generate predictions. The hyperparamerter $\mathrm{lb}$ is set to $T$. 
		\item DySAT~\cite{sankar_DySAT_2020}: It is a graph embedding model based on joint self-attention along the two dimensions of structural neighborhood and temporal dynamics. 
	\end{enumerate}
	
	\begin{table*}[t]
		\caption{Results of prediction performance in MAN and EEC. }
		\footnotesize
		\renewcommand\arraystretch{1.1}
		\begin{center}
			\begin{tabular}{ccccc}
				\toprule
				\multirow{2}[4]{*}{Method} & \multicolumn{2}{c}{MAN } & \multicolumn{2}{c}{EEC} \\
				\cmidrule{2-5}          & AUC & GMAUC & AUC & GMAUC \\
				\midrule
				TNE & $70.80 \pm 1.1$ & $77.15 \pm 1.4$ & $67.75 \pm 1.0$ & $69.87 \pm 1.1$ \\
				GC-LSTM & $72.17 \pm 0.7$ & $73.03 \pm 0.6$ & $66.72 \pm 0.7$ & $69.51 \pm 1.2$ \\
				EvolveGCN & $78.81 \pm 0.4$ & $82.53 \pm 0.3$ & $70.10 \pm 0.7$ & $72.53 \pm 0.9$ \\
				dyngraph2vec & $76.60 \pm 0.4$ & $80.16 \pm 0.4$ & $68.83 \pm 0.8$ & $70.35 \pm 1.1$ \\
				DySAT & $\mathbf{81.20} \pm 0.2$ & $84.05 \pm 0.3$ & $82.03 \pm 0.3$ & $85.50 \pm 0.2$ \\
				\textbf{TSAM} & $80.87 \pm 0.2$ & $\mathbf{84.53} \pm 0.3$ & $\mathbf{84.21} \pm 0.2$ & $\mathbf{86.75} \pm 0.2$ \\
				\bottomrule
			\end{tabular}%
			\label{tab:result1}
		\end{center}
	\end{table*}
	
	\begin{table*}[t]
		\caption{Results of prediction performance in UCI and LEM. }
		\footnotesize
		\renewcommand\arraystretch{1.1}
		\begin{center}
			\begin{tabular}{ccccc}
				\toprule
				\multirow{2}[4]{*}{Method} & \multicolumn{2}{c}{UCI} & \multicolumn{2}{c}{LEM} \\
				\cmidrule{2-5}          & AUC & GMAUC & AUC & GMAUC \\
				\midrule
				TNE & $67.11 \pm 0.7$ & $69.54 \pm 0.7$ & $76.20 \pm 0.9$ & $78.51 \pm 1.1$ \\
				GC-LSTM & $67.54 \pm 0.5$ & $70.81 \pm 0.7$ & $79.52 \pm 0.7$ & $79.90 \pm 0.7$ \\
				EvolveGCN & $69.53 \pm 0.5$ & $72.01 \pm 0.8$ & $80.71 \pm 0.4$ & $83.55 \pm 0.6$ \\
				dyngraph2vec & $75.51 \pm 0.4$ & $78.14 \pm 0.5$ & $89.15 \pm 0.4$ & $91.00 \pm 0.3$ \\
				DySAT & $79.87 \pm 0.2$ & $83.86 \pm 0.2$ & $88.52 \pm 0.3$ & $90.49 \pm 0.3$ \\
				\textbf{TSAM} & $\mathbf{81.15} \pm 0.2$ & $\mathbf{84.91} \pm 0.1$ & $\mathbf{90.81} \pm 0.3$ & $\mathbf{91.90} \pm 0.2$ \\
				\bottomrule
			\end{tabular}%
			\label{tab:result2}
		\end{center}
	\end{table*}
	
	Notice that since we are dealing with directed networks, spectral domain convolution cannot perform on the asymmetric adjacency matrices because the asymmetric laplace matrices are not decomposable. Therefore, we replace all GCN units with GraphSAGE in GC-LSTM and EvolveGCN to perform spacial domain convolution on directed networks. The mean aggregator function is utilized in GraphSAGE to aggregate neighborhood features. Moreover, since TNE, EvolveGCN and DySAT are graph embedding models which learn node embedding $\mathbf{e}_{t}$ from the evolving network $G_{t-T}^t$, we define similarity score matrix $\mathbf{S}_{t+1} = \mathbf{e}_t \cdot \mathbf{e}_t^\mathrm{T}$ to use them for temporal link prediction task. 	
	% Parameter
	In TSAM, we adopt four transformations $\{\mathbf{C}^{M_1}, \mathbf{C}^{M_2}, \mathbf{C}^{M_3}, \mathbf{C}^{M_4}\}$ listed in~(\ref{eq:transexp}) as additional feature inputs. Other parameter settings in four networks are presented in Table~\ref{tab:parameter}. 

	\begin{figure}[t]
		\centerline{\includegraphics[width=14 cm]{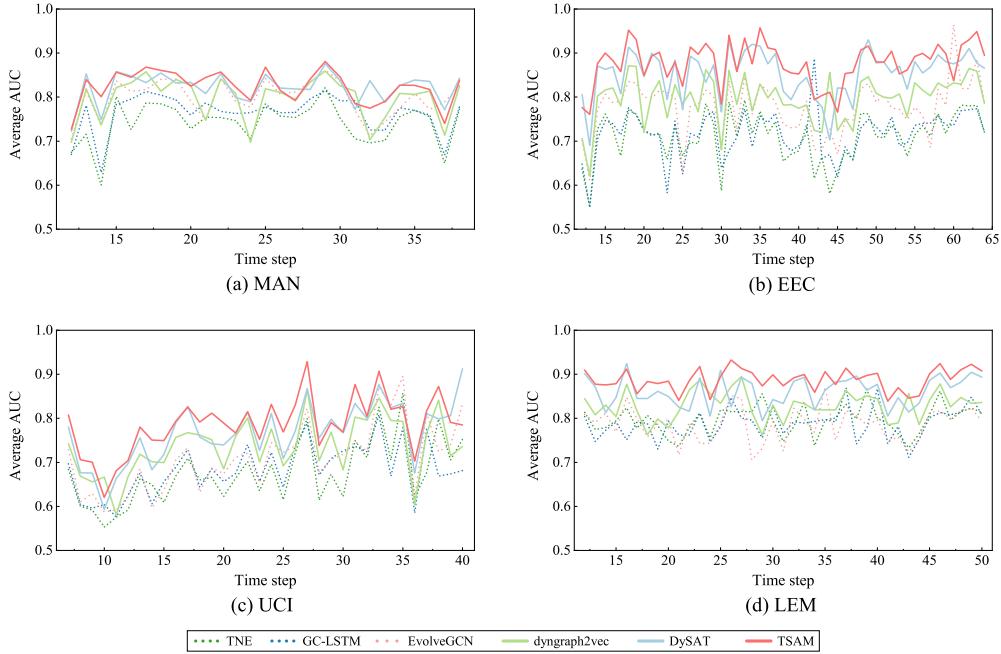}}
		\caption{Average AUC of TSAM and baselines in four networks at each time step. Each value is the average of $20$ independent experiments. }
		\label{fig:timeauc}
	\end{figure}

	Table~\ref{tab:result1} and Table~\ref{tab:result2} presents the prediction performance of TSAM and baselines in four networks. From the results we can observe consistent gains of $1-2\%$ AUC in comparison to the best baseline in EEC, UCI and LEM. In MAN, the AUC of TSAM is slightly lower than DySAT, but TSAM get higher GMAUC than DySAT, which indicates a better performance of predicting both added and removed links. It is also clear that the performance of TSAM is more stable in four networks reflected by the smaller standard deviation. We also notice that compared with dyngraph2vec and GC-LSTM, our TSAM model and DySAT are able to achieve obvious improvements on both AUC and GMAUC, which indicates the effectiveness of self-attention mechanism on temporal link prediction task.   

	Fig.~\ref{fig:timeauc} presents the average AUC at each time step in four networks. We find that the performance of TSAM is relatively better and more stable than baselines. The superiority of TSAM is obvious in comparison with TNE and GC-LSTM, whose AUC curves show rapid drops at certain time steps. 
	
	\begin{table*}[t]
		\caption{Comparison on the influence of feature fusion in MAN and EEC. }
		\footnotesize
		\renewcommand\arraystretch{1.2}
		\begin{center}
			\begin{tabular}{ccccc}
				\toprule
				\multirow{2}[4]{*}{Method} & \multicolumn{2}{c}{MAN } & \multicolumn{2}{c}{EEC} \\
				\cmidrule{2-5}          & AUC & GMAUC & AUC & GMAUC \\
				\midrule
				No feature & $78.11 \pm 0.3$ & $82.31 \pm 0.3$ & $80.15 \pm 0.3$ & $81.24 \pm 0.3$ \\
				$\{\mathbf{C}^{M_1}\}$ & $80.19 \pm 0.2$ & $81.95 \pm 0.3$ & $82.60 \pm 0.3$ & $83.83 \pm 0.2$ \\
				$\{\mathbf{C}^{M_1}, \mathbf{C}^{M_2}, \mathbf{C}^{M_3}, \mathbf{C}^{M_4}\}$ & $\mathbf{80.87} \pm 0.2$ & $\mathbf{84.53} \pm 0.3$ & $\mathbf{84.21} \pm 0.2$ & $\mathbf{86.75} \pm 0.2$ \\
				\bottomrule
			\end{tabular}%
			\label{tab:result3}
		\end{center}
	\end{table*}
	
	\begin{table*}[t]
		\caption{Comparison on the influence of feature fusion in UCI and LEM. }
		\footnotesize
		\renewcommand\arraystretch{1.2}
		\begin{center}
			\begin{tabular}{ccccc}
				\toprule
				\multirow{2}[4]{*}{Method} & \multicolumn{2}{c}{UCI} & \multicolumn{2}{c}{LEM} \\
				\cmidrule{2-5}          & AUC & GMAUC & AUC & GMAUC \\
				\midrule
				No feature & $80.81 \pm 0.2$ & $82.53 \pm 0.2$ & $89.14 \pm 0.3$ & $91.12 \pm 0.3$ \\
				$\{\mathbf{C}^{M_1}\}$ & $79.90 \pm 0.2$ & $81.97 \pm 0.2$ & $89.57 \pm 0.3$ & $90.26 \pm 0.2$ \\
				$\{\mathbf{C}^{M_1}, \mathbf{C}^{M_2}, \mathbf{C}^{M_3}, \mathbf{C}^{M_4}\}$ & $\mathbf{81.15} \pm 0.2$ & $\mathbf{84.91} \pm 0.1$ & $\mathbf{90.81} \pm 0.3$ & $\mathbf{91.90} \pm 0.2$ \\
				\bottomrule
			\end{tabular}%
			\label{tab:result4}
		\end{center}
	\end{table*}
	
	\subsection{Influence of feature fusion}\label{fusion}

	An important improvement in TSAM compared with other baselines is the feature fusion block, which uses matrix transformation to capture motif features in directed networks. Here we analyze the influence of feature fusion block with different choices of additional feature. Table~\ref{tab:result3} and Table~\ref{tab:result4} present the performance of TSAM under three circumstances: 1) use only the output of GAT without additional features as the input of GRU, 2) use $\mathbf{C}^{M_1}$ as additional feature, 3) use matrices listed in Eq.~(\ref{eq:transform}) as additional feature. We adopt parameters under the best performance for each case. From the results we find that without additional feature, TSAM achieves lower AUC and GMAUC than DySAT in MAN and EEC. However, when $\mathbf{C}^{M_1}$ is adopted, an obvious improvement can be observed in four networks. Since $\mathbf{C}^{M_1}$ captures the local motifs of $u \to t \to v$, it indicates that taking into account the motif structure in directed networks can lead to improvement on prediction performance. We also find that with more additional features added, the performance of TSAM gets better, bringing along more computational costs at the same time. 
	
	\subsection{Influence of attention heads}\label{attnhead}
	
\begin{figure}[t]
		\centerline{\includegraphics[width=14 cm]{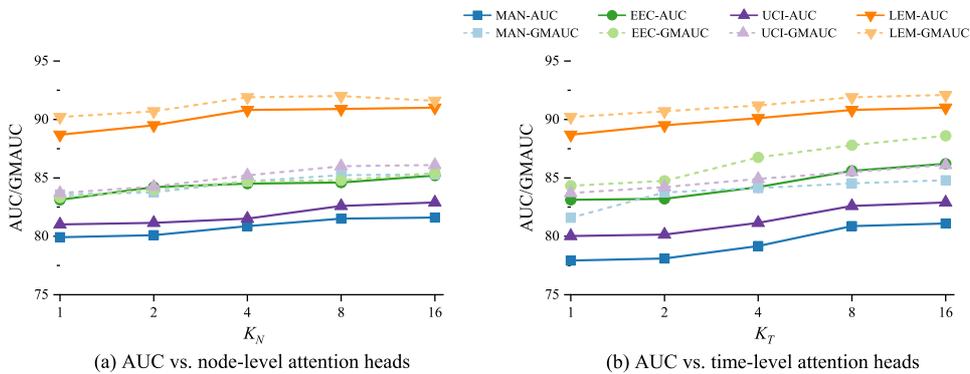}}
		\caption{Performance comparison on the number of attention heads in node-level and time-level self-attention blocks. }
		\label{fig:nattn}
	\end{figure}

	Another important hyperparameter of TSAM is the number of attention heads in self-attention layers. Here we respectively analyze the influence of attention heads on the performance of TSAM. Fig.~\ref{fig:nattn} presents the performance of TSAM when we independently vary the number of attention heads in node-level and time-level self-attention blocks within range $\{1,2,4,8,16\}$. In the results we find that more attention heads leads to better performance on both AUC and GMAUC. When $K_N \geq 4$ and $K_T \geq 8$, the performance tend to be stable. It indicates that it is efficient to capture latent features when using $4$ node-level attention heads and $8$ time-level attention heads. 
	
	\section{Conclusions}\label{sec:conclusion}
	Predicting the connectivity and direction of links in temporal networks is both meaningful and challenging. In this paper we propose a temporal link prediction model for directed networks based on graph neural networks and self-attention mechanism. The basic architecture of our model is an autoencoder. In the encoder, local structural features of each snapshot are drawn by the GAT layer and several GCLs. A GRU hidden layer then captures the temporal variations of the snapshot sequence. We use a time-level self-attention layer to differentiate the effect of each snapshot. In the decoder, a full-connected layer transforms the learned embedding into predicted adjacency matrix. Experimental results on realistic networks prove the effectiveness of our model in comparison with baselines. 
	
	In our future works, we will focus on the prediction of weighted links in directed networks. A possible direction of extending TSAM to solve weight prediction problems is leveraging the structure of generative adversarial network to refine the prediction accuracy. 
	
	\section*{Acknowledgment}
	This research was funded by the Foundation for Innovative Research Groups of the National Natural Science Foundation of China (Grant No. 61521003), and National Natural Science Foundation of China (Grant No. 61803384). 
	
	\bibliographystyle{unsrt}
	\bibliography{bibfile}

\begin{thebibliography}{10}

\bibitem{kumar_Structure_2006}
Ravi Kumar, Jasmine Novak, and Andrew Tomkins.
\newblock Structure and evolution of online social networks.
\newblock In {\em Proceedings of the 12th {ACM} {SIGKDD} international
  conference on {Knowledge} discovery and data mining - {KDD} '06}, pages
  611--617, Philadelphia, PA, USA, 2006. ACM Press.

\bibitem{lei_GCNGAN_2019}
Kai Lei, Meng Qin, Bo~Bai, Gong Zhang, and Min Yang.
\newblock {GCN}-{GAN}: {A} non-linear temporal link prediction model for
  weighted dynamic networks.
\newblock In {\em {IEEE} {INFOCOM} 2019 - {IEEE} {Conference} on {Computer}
  {Communications}}, pages 388--396, April 2019.

\bibitem{Lu_LinkPrediction_2011}
Linyuan Lü and Tao Zhou.
\newblock Link prediction in complex networks: {A} survey.
\newblock {\em Phys. A}, 390(6):1150--1170, March 2011.

\bibitem{Tylenda_Timeaware_2009}
Tomasz Tylenda, Ralitsa Angelova, and Srikanta Bedathur.
\newblock Towards time-aware link prediction in evolving social networks.
\newblock In {\em Proceedings of the 3rd {Workshop} on {Social} {Network}
  {Mining} and {Analysis} - {SNA}-{KDD} '09}, pages 1--10, Paris, France, 2009.
  ACM Press.

\bibitem{Liben_Linkprediction_2007}
David Liben-Nowell and Jon Kleinberg.
\newblock The link-prediction problem for social networks.
\newblock {\em Journal of the American Society for Information Science and
  Technology}, 58(7):1019--1031, May 2007.

\bibitem{Kipf_Semisupervised_2017}
Thomas~N. Kipf and Max Welling.
\newblock Semi-supervised classification with graph convolutional networks.
\newblock In {\em International {Conference} on {Learning} {Representations}},
  April 2017.

\bibitem{sankar_DySAT_2020}
Aravind Sankar, Yanhong Wu, Liang Gou, Wei Zhang, and Hao Yang.
\newblock {DySAT}: {Deep} {Neural} {Representation} {Learning} on {Dynamic}
  {Graphs} via {Self}-{Attention} {Networks}.
\newblock In {\em Proceedings of the 13th {International} {Conference} on {Web}
  {Search} and {Data} {Mining}}, {WSDM} '20, pages 519--527, Houston, TX, USA,
  January 2020. Association for Computing Machinery.

\bibitem{Shang_RoleDirect_2017}
Ke-ke Shang, Michael Small, Xiao-ke Xu, and Wei-sheng Yan.
\newblock The role of direct links for link prediction in evolving networks.
\newblock {\em EPL}, 117(2):28002, January 2017.

\bibitem{Zhou_Dynamic_2018}
Lekui Zhou, Yang Yang, Xiang Ren, Fei Wu, and Yueting Zhuang.
\newblock Dynamic network embedding by modeling triadic closure process.
\newblock In {\em Thirty-second {AAAI} conference on artificial intelligence},
  2018.

\bibitem{Goyal_DynGEM_2018}
Palash Goyal, Nitin Kamra, Xinran He, and Yan Liu.
\newblock {DynGEM}: {Deep} {Embedding} {Method} for {Dynamic} {Graphs}.
\newblock {\em arxiv}, May 2018.

\bibitem{Chen_GCLSTM_2018}
Jinyin Chen, Xuanheng Xu, Yangyang Wu, and Haibin Zheng.
\newblock {GC}-{LSTM}: {Graph} {Convolution} {Embedded} {LSTM} for {Dynamic}
  {Link} {Prediction}.
\newblock {\em arxiv}, December 2018.

\bibitem{Pareja_EvolveGCN_2019}
Aldo Pareja, Giacomo Domeniconi, Jie Chen, Tengfei Ma, Toyotaro Suzumura,
  Hiroki Kanezashi, Tim Kaler, Tao~B. Schardl, and Charles~E. Leiserson.
\newblock {EvolveGCN}: {Evolving} {Graph} {Convolutional} {Networks} for
  {Dynamic} {Graphs}.
\newblock {\em arxiv}, February 2019.

\bibitem{Jawed_TimeFrame_2015}
Mujtaba Jawed, Mehmet Kaya, and Reda Alhajj.
\newblock Time frame based link prediction in directed citation networks.
\newblock In {\em Proceedings of the 2015 {IEEE}/{ACM} {International}
  {Conference} on {Advances} in {Social} {Networks} {Analysis} and {Mining}
  2015 - {ASONAM} '15}, pages 1162--1168, Paris, France, 2015. ACM Press.

\bibitem{Butun_Extension_2018}
Ertan Bütün, Mehmet Kaya, and Reda Alhajj.
\newblock Extension of neighbor-based link prediction methods for directed,
  weighted and temporal social networks.
\newblock {\em Information Sciences}, 463-464:152--165, October 2018.

\bibitem{hamilton_Inductive_2017}
William~L. Hamilton, Rex Ying, and Jure Leskovec.
\newblock Inductive representation learning on large graphs.
\newblock {\em arxiv}, June 2017.

\bibitem{Dareddy_Motif2vec_2019}
Manoj~Reddy Dareddy, Mahashweta Das, and Hao Yang.
\newblock Motif2vec: {Motif} aware node representation learning for
  heterogeneous networks.
\newblock August 2019.

\bibitem{Kefato_WhichWay_2020}
Zekarias~T. Kefato, Nasrullah Sheikh, and Alberto Montresor.
\newblock Which way? {Direction}-{Aware} {Attributed} {Graph} {Embedding}.
\newblock {\em arxiv}, January 2020.

\bibitem{Sun_ATPDirected_2018}
Jiankai Sun, Bortik Bandyopadhyay, Armin Bashizade, Jiongqian Liang,
  P.~Sadayappan, and Srinivasan Parthasarathy.
\newblock {ATP}: {Directed} {Graph} {Embedding} with {Asymmetric}
  {Transitivity} {Preservation}.
\newblock {\em arxiv}, November 2018.

\bibitem{Monti_MotifNet_2018}
Federico Monti, Karl Otness, and Michael~M. Bronstein.
\newblock {MotifNet}: {A} motif-based graph convolutional network for directed
  graphs.
\newblock In {\em 2018 {IEEE} {Data} {Science} {Workshop} ({DSW})}, pages
  225--228, Lausanne, Switzerland, June 2018. IEEE.

\bibitem{Velickovic_GraphAttention_2017}
Petar Veličković, Guillem Cucurull, Arantxa Casanova, Adriana Romero, Pietro
  Liò, and Yoshua Bengio.
\newblock Graph attention networks.
\newblock {\em arxiv}, October 2017.

\bibitem{Gu_LinkPrediction_2019}
Weiwei Gu, Fei Gao, Xiaodan Lou, and Jiang Zhang.
\newblock Link prediction via graph attention network.
\newblock {\em arXiv}, October 2019.

\bibitem{Zhang_GaANGated_2018}
Jiani Zhang, Xingjian Shi, Junyuan Xie, Hao Ma, Irwin King, and Dit-Yan Yeung.
\newblock {GaAN}: {Gated} {Attention} {Networks} for {Learning} on {Large} and
  {Spatiotemporal} {Graphs}.
\newblock {\em arXiv}, March 2018.

\bibitem{Bianconi_LocalStructure_2008}
Ginestra Bianconi, Natali Gulbahce, and Adilson~E. Motter.
\newblock Local structure of directed networks.
\newblock {\em Phys. Rev. Lett.}, 100(11), March 2008.

\bibitem{vaswani_Attention_2017}
Ashish Vaswani, Noam Shazeer, Niki Parmar, Jakob Uszkoreit, Llion Jones,
  Aidan~N. Gomez, Lukasz Kaiser, and Illia Polosukhin.
\newblock Attention {Is} {All} {You} {Need}.
\newblock {\em arxiv}, June 2017.

\bibitem{voita_Analyzing_2019}
Elena Voita, David Talbot, Fedor Moiseev, Rico Sennrich, and Ivan Titov.
\newblock Analyzing {Multi}-{Head} {Self}-{Attention}: {Specialized} {Heads}
  {Do} the {Heavy} {Lifting}, the {Rest} {Can} {Be} {Pruned}.
\newblock {\em arxiv}, May 2019.

\bibitem{Wang_Structure_2016}
Daixin Wang, Peng Cui, and Wenwu Zhu.
\newblock Structural {Deep} {Network} {Embedding}.
\newblock In {\em Proceedings of the 22nd {ACM} {SIGKDD} {International}
  {Conference} on {Knowledge} {Discovery} and {Data} {Mining}}, {KDD} '16,
  pages 1225--1234, San Francisco, California, USA, August 2016. Association
  for Computing Machinery.

\bibitem{Kunegis_KONECT_2013}
Jérôme Kunegis.
\newblock {KONECT}: the {Koblenz} network collection.
\newblock In {\em Proceedings of the 22nd {International} {Conference} on
  {World} {Wide} {Web} - {WWW} '13 {Companion}}, pages 1343--1350, Rio de
  Janeiro, Brazil, 2013. ACM Press.

\bibitem{Paranjape_Motifs_2017}
Ashwin Paranjape, Austin~R. Benson, and Jure Leskovec.
\newblock Motifs in {Temporal} {Networks}.
\newblock In {\em Proceedings of the {Tenth} {ACM} {International} {Conference}
  on {Web} {Search} and {Data} {Mining} - {WSDM} '17}, pages 601--610,
  Cambridge, United Kingdom, 2017. ACM Press.

\bibitem{Panzarasa_Patterns_2009}
Pietro Panzarasa, Tore Opsahl, and Kathleen~M. Carley.
\newblock Patterns and dynamics of users’ behavior and interaction: Network
  analysis of an online community.
\newblock {\em J. Am. Soc. Inf. Sci. Technol.}, 60(5):911–932, May 2009.

\bibitem{Leskovec_SNAP_2014}
Jure Leskovec and Andrej Krevl.
\newblock {SNAP} {Datasets}: {Standford} large network dataset collection,
  January 2014.

\bibitem{Junuthula_Evaluating_2016}
R.~R. {Junuthula}, K.~S. {Xu}, and V.~K. {Devabhaktuni}.
\newblock Evaluating link prediction accuracy in dynamic networks with added
  and removed edges.
\newblock In {\em 2016 IEEE International Conferences on Big Data and Cloud
  Computing (BDCloud), Social Computing and Networking (SocialCom), Sustainable
  Computing and Communications (SustainCom) (BDCloud-SocialCom-SustainCom)},
  pages 377--384, 2016.

\bibitem{Zhu_Scalable_2016}
Linhong Zhu, Dong Guo, Junming Yin, Greg~Ver Steeg, and Aram Galstyan.
\newblock Scalable {Temporal} {Latent} {Space} {Inference} for {Link}
  {Prediction} in {Dynamic} {Social} {Networks}.
\newblock {\em IEEE Trans. Knowl. Data Eng.}, 28(10):2765--2777, October 2016.

\bibitem{Goyal_Dyngraph2vec_2018}
Palash Goyal, Sujit~Rokka Chhetri, and Arquimedes Canedo.
\newblock dyngraph2vec: {Capturing} {Network} {Dynamics} using {Dynamic}
  {Graph} {Representation} {Learning}.
\newblock {\em arxiv}, September 2018.

\end{thebibliography}
	
\end{document}